\documentclass[10pt,a4paper]{article} 
\usepackage[nochapters, subfig, listings]{classicthesis} 

\usepackage[english]{babel}
\usepackage{lipsum}
\usepackage[smaller]{acronym}

\usepackage{url}
\usepackage{dsfont}
\usepackage{mathcmd}
\usepackage{enumerate}
\usepackage{amsfonts}
\usepackage{amsmath}
\usepackage{amsthm}
\usepackage{amssymb} 
\usepackage{graphicx}
\usepackage{nicefrac}
\usepackage{enumitem}
\usepackage{subfig}
\usepackage[T1]{fontenc} 
\usepackage{microtype}
\usepackage[affil-it]{authblk}
\newcounter{thmctr}
\newenvironment{thm}{
   \par\bigskip\noindent
   \refstepcounter{thmctr}
   \spacedlowsmallcaps{Theorem \thethmctr}
   \begin{it}
   }
   {\end{it}
   \par\medskip}  
\newenvironment{prp}{
   \par\medskip\noindent
   \refstepcounter{thmctr}
   \spacedlowsmallcaps{Proposition \thethmctr}
   \begin{it}
   }
   {\end{it}
   \par\medskip}  
\theoremstyle{remark}
\newtheorem{rmk}{Remark}
\newtheorem{dfn}{Definition}
\numberwithin{equation}{section}

\begin{document}
    \title{\rmfamily\normalfont\spacedallcaps{Equilibrium and non-equilibrium Ising models by means of PCA}}

    \author[1,2]{Carlo Lancia}
    \author[1]{Benedetto Scoppola}
    \affil[1]{Universit\`a di Roma `Tor Vergata'}
    \affil[2]{TU/e, Technische Universiteit Eindhoven}
    
    \maketitle

    \acrodef{PCA}{Probabilistic Cellular Automaton}
    \acrodef{CUDA}{Compute Unified Device Architecture}
    \acrodef{GPU}{Graphics Processing Unit}
    \acrodef{CPU}{Central Processing Unit}
    \acrodef{MC}{Markov chain}
    \acrodef{CA}{Cellular Automata}
    
    \begin{abstract}
\noindent
We propose a unified approach to reversible and irreversible \acs{PCA} dynamics,
and we show that in the case of $1$D and $2$D nearest neighbour Ising systems
with periodic boundary conditions we are able to compute the stationary
measure of the dynamics also when the latter is irreversible
We also show how, according to~\cite{dai2012sampling}, the stationary measure is
very close to the Gibbs for a suitable choice of the parameters of the
\acs{PCA} dynamics, both in the reversible and in the irreversible cases. We discuss 
some numerical aspects regarding this topic, including a possible
parallel implementation.
\end{abstract}

\section{Introduction}
\label{sec:introduction}

In this paper we propose a connection between two different subjects
that have been quite studied over the last two decades, that is, the
general study of non-equilibrium statistical mechanics and the
description of equilibrium statistical mechanics with the specific use
of a \ac{PCA}. 
Starting from a series of seminal papers,
e.g.~\cite{goldstein1989pca,lebowitz1990statistical,maes1993interacting}, a
certain effort has been spent in order to give a dynamical description
of the equilibrium statistical mechanics by means of \acp{PCA}.
A \ac{PCA} is a discrete-time \ac{MC} on a product space $S^V$
such that the transition probability $P(\sigma,\tau)$ is a product
measure
\begin{equation}
  \label{eq:1}
  P(\sigma,\tau) = \prod_{i \in V} p_i(\tau_i \,|\,\sigma) \;,
\end{equation}
where $p_i(\tau_i \,|\,\sigma)$
is a probability on $S$ for all $i \in V$ and $\sigma \in S^V$.  
Given a probability measure $\mu$ on $S^V$, the question is whether a \ac{PCA}
may be constructed whose stationary measure is $\mu$. 
In the context of the equilibrium statistical mechanics
we are obviously interested to the case in which $\mu$ is a Gibbs measure.
It is well known that such a problem is rather difficult. 
For instance, in \cite{kozlov1980reversible} it is shown that no
\ac{PCA} can be designed in such a way to be reversible with respect to the
stationary distribution of a $2$D Ising model. 
In~\cite{lebowitz1990statistical} and~\cite{cirillo2003metastability} 
a \ac{PCA} is introduced 
whose invariant, reversible measure $\pi$
is related to the Ising model as follows. 
The projection of $\pi$ to
the {\em even} sites, i.e., those $(i,j) \in \mathbb{Z}^2$  
with $i+j$ even, coincides with the same projection 
of the Ising model, and the same holds for odd sites. 
However, opposite to the Ising model, spins at
even sites are independent under the measure $\pi$
of those at odd sites.
Recently, one of the authors has proved in~\cite{dai2012sampling} that
the \ac{PCA} from~\cite{cirillo2003metastability} can be simply modified in order 
to provide a way to approximately sample from a Gibbs 
measure.
The sense of this statement will be clarified in Section~\ref{sec:revers-irrev-pca}.
This result was achieved following some ideas first introduced
in~\cite{iovanella2007some} and later exploited in~\cite{scoppola2011exact}
to provide an easy example of cutoff.

On the other hand, one of the main subjects of the recent research in
statistical mechanics has been the study of non-equilibrium
statistical mechanics, in particular when a stationary state can be defined. 
Several different approaches have been proposed, see
e.g.~\cite{gallavotti1995dynamical,gallavotti1995dynamical2},~\cite{jona2010fluctuations} 
and the references therein.   
In contrast with the case of equilibrium statistical mechanics,
where a well defined recipe, the Gibbs measure, is always available,
in the non-equilibrium case the subject is far from a complete and
general understanding. The study of the stationary measure in the
non-equilibrium context can not leave aside the dynamics of the
process whereas the Gibbs recipe has exactly this advantage. 

In this paper we propose, starting from the results in~\cite{dai2012sampling},  a
unified description of equilibrium and non-equilibrium statistical
mechanics in terms of \ac{PCA}. 
It turns out that such a description gives to the chance to compute explicitly the
stationary state of a set of the statistical mechanics models,
including the nearest neighbour Ising model in 1 and 2
dimensions. 
An extension of the class of system for which this approach give
rise to explicit computations seems very likely with the use of
relatively easy arguments. 

The paper is organised as follows. 
In Section~\ref{sec:revers-irrev-pca} we will define, according
to~\cite{dai2012sampling},  
a class of reversible and irreversible \ac{PCA} dynamics. We will prove
that if a \emph{weak balance condition}
is verified then we are able to explicitly write the unique stationary
measure of the chain.
In Section~\ref{sec:glob-balanc-ising} we show that the Ising systems
introduced in Section~\ref{sec:revers-irrev-pca} are weakly balanced
whenever periodic boundary conditions are imposed. 
Finally, in Section~\ref{sec:numer-results:-ising} we present some
numerical/simulative aspects of \acp{PCA}.
In particular, we show that in the irreversible case a stationary current is present in the system,
and under suitable conditions what we call \emph{Ising waves} arise.

\section{Reversible and irreversible PCA Dynamics}
\label{sec:revers-irrev-pca}

Let us consider a {\it two-body spin system}, defined by an Hamiltonian of the form
\begin{equation}
    \label{twob}
    H(\sigma) = -\sum_{(x,y)} J_{\{x,y\}}\sigma_x\,\sigma_y \;,
  \end{equation}
where $\Lambda$ is a finite set, $\sigma\in{\cal
  X}=\{-1,+1\}^\Lambda$, and the sum in~\eqref{twob} is extended 
to the unordered pairs $\{x,y\}$ with $x,y\in\Lambda$.
The Gibbs measure associated to the Hamiltonian~\eqref{twob} is then
\begin{equation}
    \label{gibbs}
    \pi_G(\sigma)=\frac {e^{-H(\sigma)}}{Z} \;,
  \end{equation}
where $Z=\sum_\sigma e^{-H(\sigma)}$.
It is possible to construct Markovian, ergodic dynamics having stationary measure
is precisely~$\pi_G(\sigma)$.
The long run behaviour of the chain can then be
used to sample from~$\pi_G(\sigma)$.
Those Markovian
dynamics are typically based
on single spin-flip transition probabilities, see
for instance~\cite{pierre1999markov,levin2009markov} 
and the references therein. 

One of the authors has recently proved in~\cite{dai2012sampling} that
an approximate sampling of Gibbs measure~\eqref{gibbs} can be actually
achieved also by means of a reversible \ac{PCA} dynamics.
Consider the following Hamiltonian, defined on pairs of configurations $(\sigma,\tau)$,
\begin{equation}\label{revers1}
  H(\sigma, \tau) = -\sum_{(x,y)} J_{x,y}\,\sigma_x\,\tau_y
    - q\, \sum_x\sigma_x\,\tau_x \;,
    \end{equation}
where
\begin{equation}
  \label{eq:5}
  J_{x,y}=J_{y,x}=\frac{J_{\{x,y\}}}{2} \;,\quad \forall\, x,y \in \Lambda\,.
\end{equation}
The last term in the Hamiltonian~\eqref{revers1}, proportional to the parameter $q>0$, represents an inertial term, 
i.e., the tendency of the system to remain in the current state $\sigma$.
We next define a homogeneous \ac{MC} whose transition probabilities are
\begin{equation}
\label{eq:6}
      P(\sigma, \tau) =
      \frac{e^{-H(\sigma, \tau)}}{Z_\sigma}\;,
\end{equation}
where
\begin{equation}
\label{eq:11}
Z_\sigma=\sum_\tau e^{-H(\sigma,\tau)}\,.
\end{equation}
From~\eqref{eq:5}
$H(\sigma,\sigma^\prime)=H(\sigma^\prime,\sigma)$,
thus the Hamiltonian~\eqref{revers1} is symmetric for the exchange $\sigma\leftrightarrow\sigma^\prime$.
It is then immediate to see that such a \ac{MC} is ergodic and reversible with respect to the measure
  \begin{equation}
   \label{eq:stationary1}
    \pi(\sigma) = \frac{Z_\sigma}{Z}\;,
  \end{equation}
where $Z=\sum_{\sigma}Z_\sigma$. 
The following result holds
\begin{thm}{\cite{dai2012sampling}}
\label{thm-dss}
If $e^{-2q} = o\left(|\Lambda|^{\frac{1}{2}}\right)$ as
$|\Lambda| \to \infty$ then
\begin{equation}
  \label{eq:4}
  \lim_{\Lambda\to\infty} d_{\text{\tiny{TV}}}\left( \pi, \pi_G
  \right) = 0
\end{equation}
where $d_{\text{\small{TV}}}(\cdot,\cdot)$ is the usual
total-variation distance and $\pi_G$ is
given by~\eqref{twob}--\eqref{gibbs}.
\end{thm}

\begin{rmk}
  \label{rmk:beethoven_no5}
  The long run behaviour of the collective dynamics defined
  by~\eqref{eq:6} can be used to perform an approximate sampling of a
  Gibbs distribution in the sense of~\eqref{eq:4}.  Note that
  in~\cite{dai2012sampling} the limit~\eqref{eq:4} is obtained without
  making any use of reversibility.
\end{rmk}
Having in mind a generalisation of this picture to irreversible \acp{MC},
let us define a non-symmetric Hamiltonian on pairs of configurations
by simply taking~\eqref{revers1}, where now $J_{x,y}\ne J_{y,x}$.
\begin{dfn}
  \label{defn:eroica}
  Whenever $J_{x,y}\ne 0\Rightarrow J_{y,x}=0$ we will say that the
  spin system~\eqref{revers1} is {\it completely asymmetric}. 
\end{dfn}
The new chain will evolve according to~\eqref{eq:6}. 
Due to the lack of symmetry of the interaction $J_{x,y}$,
the \ac{MC} is no longer reversible with respect to~\eqref{eq:stationary1} 
because $H(\sigma,\tau) \ne H(\tau,\sigma)$ and the detailed balance
condition is not satisfied.
However, we can still consider $\pi(\sigma)$ given by~\eqref{eq:stationary1}
and compute
$\sum_\sigma\pi(\sigma)P(\sigma,\tau)$. This yields
\begin{equation}
  \label{statirr}
  \sum_\sigma\pi(\sigma)P(\sigma,\tau) =
  \sum_\sigma\frac{Z_\sigma}{Z}\frac{e^{-H(\sigma,\tau)}}{Z_\sigma}=
  \sum_\sigma\frac{e^{-H(\sigma,\tau)}}{Z}\,.
\end{equation}

\begin{prp}
  \label{prop-WBC}\acrodef{WBC}{Weak Balance Condition}
  Consider the following \ac{MC}
  \begin{align}
    \label{transirr}
    P(\sigma,\tau) &= \frac{e^{-H(\sigma,\tau)}}{Z_\sigma} \;,\\
    \label{irrevers1}
    H(\sigma,\tau) &= -\sum_{(x,y)} J_{x,y}\sigma_x\,\tau_y
    - q\, \sum_x\sigma_x\,\tau_x \;,\\
    \label{eq:2}
    Z_\sigma &= \sum_\tau e^{-H(\sigma,\tau)}\;,
 \end{align}
 where no assumption is made on $J_{x,y}$.
 Suppose the following \emph{\ac{WBC}} holds
  \begin{equation}
    \label{eq:weakreversibility1}
    \sum_{\tau}e^{-H(\sigma,\tau)} =
    \sum_{\tau}e^{-H(\tau,\sigma)} \;,\quad \forall\;
    \sigma \in {\cal X}\,.
  \end{equation}
  Then the probability distribution
  \begin{equation}
   \label{eq:stationary2}
    \pi(\sigma) = \frac{Z_\sigma}{Z} \quad \text{with} \quad Z=\sum_{\sigma}Z_\sigma \;,
  \end{equation}
  is the stationary distribution of the chain.
\end{prp}
\begin{proof}
  By~\eqref{statirr} and~\eqref{eq:weakreversibility1},
  \begin{equation}
    \label{eq:3}
    \sum_\sigma\pi(\sigma)P(\sigma,\tau) = \sum_{\tau}
    \frac{e^{-H(\tau,\sigma)}}{Z} = \frac{Z_\tau}{Z} = \pi(\tau)\,.
  \end{equation}
  Therefore~\eqref{eq:stationary2} is stationary.
\end{proof}

\begin{rmk}
  The \ac{WBC}~\eqref{eq:weakreversibility1} determines a class of
  \acp{MC} for which the stationary distribution
  is~\eqref{eq:stationary2}, some examples within this class are
  presented in Section~\ref{sec:glob-balanc-ising}.
  The fulfilment of the \ac{WBC} is indeed a weaker requirement
  than the detailed balance principle, which holds for the reversible
  dynamics due to~\eqref{eq:5}.
\end{rmk}
We end the present section with a couple of important remarks
\begin{rmk}
  Given a function $g: {\cal X} \times {\cal X} \to \mathbb{R}$, 
  its expectation with respect to the stationary measure 
  over pairs of subsequent configurations is
  \begin{equation}
    \label{twostep}
    \langle g(\sigma,\tau)\rangle=\sum_{\sigma,\tau}\pi(\sigma)P(\sigma,\tau)g(\sigma,\tau)=
    \sum_{\sigma,\tau}\frac{e^{-H(\sigma,\tau)}g(\sigma,\tau)}{Z} \,.
  \end{equation}
  Hence, the two-step stationary measure of the dynamics is Gibbsian
  in the sense of~\eqref{twostep}.
\end{rmk}
\begin{rmk}
  Given a function $g: {\cal X} \times {\cal X} \to \mathbb{R}$,
  consider
  \begin{equation}
    \label{Jdig}
    \mathcal{J}(g) =\langle g(\sigma,\tau)-g(\tau,\sigma)\rangle \,.
  \end{equation}
  Whenever $\mathcal{J}(g) \ne 0$ we see a \emph{current} of the function $g$.
  In the reversible case $\mathcal{J}(g)$ is identically zero for each
  $g$, while in the irreversible case it may be in general different
  from zero.
\end{rmk}

\section{Weakly balanced Ising irreversible systems}
\label{sec:glob-balanc-ising}

We prove the \ac{WBC}~\eqref{eq:weakreversibility1} 
for some examples of $1$D and $2$D nearest neighbour
Ising systems. 
Similar proofs can be done in more general cases.

\subsection{Weak balance for $1$D nearest neighbour Ising systems}
The Hamiltonian of the $1$D nearest neighbour Ising model is given by
 \begin{equation}
    \label{ising1D}
    H(\sigma) = -\sum_{i=1}^L J_i\,\sigma_{i}\,\sigma_{i+1} \,.
  \end{equation}
Here we assume that the strength of the interaction $J_i$ may in general depend
on the site $i$. The sum on $i$ may be performed using either periodic
boundary conditions or empty boundary condition. In the first case
sites $1$ and $L+1$ coincide (Ising system on the circle).  
In the second case $J_L=0$ (Ising system on the segment).

We can introduce a \ac{PCA} reversible dynamics for this model starting from the pair Hamiltonian
 \begin{equation}
    \label{ising1Drev}
    H(\sigma,\tau) = -\sum_{i=1}^L \left[\frac{J_i}{2}\,(\sigma_{i}\,\tau_{i+1}+\tau_{i}\,\sigma_{i+1}) + 
    q\sigma_i\tau_i\right] \;,
  \end{equation}
and then defining the \ac{PCA} dynamics with the transition probabilities
\begin{equation}\label{mark1D}
      P(\sigma,\tau) = \frac{e^{-H(\sigma,\tau)}}{Z_\sigma} \,.
\end{equation}
Both periodic and empty boundary conditions can be considered for the
sum~\eqref{ising1Drev}.
This reversible \ac{PCA} dynamics can be generalised to an irreversible case in two ways. The first
possibility is to define a completely asymmetric system by the pair Hamiltonian
\begin{equation}
    \label{ising1Dirrev}
    H(\sigma,\tau) = -\sum_{i=1}^L \left[J_i\,\sigma_{i}\,\tau_{i+1}+ 
    q\sigma_i\tau_i\right] \,.
  \end{equation}
This is a particularly simple irreversible system, and the \ac{WBC}
can be directly checked in one line
\begin{align}
  \sum_{\tau}e^{-H(\sigma,\tau)} 
 &=2^L\prod_i\cosh(J_{i-1}\sigma_{i-1}+q\sigma_{i})\nonumber\\
 &= 2^L\prod_i\cosh(J_{i-1}\sigma_{i}+q\sigma_{i-1})
  = \sum_{\tau}e^{-H(\tau,\sigma)} \;,\label{global1D}
\end{align}
where in the second equality we have used the parity of the hyperbolic cosine.
We will now compute $Z$ using the following elementary identities
\begin{align}
\cosh(a+b)&=\cosh a\cosh b + \sinh a \sinh b \;, \label{eq:9}\\
\sinh(a+b)&=\sinh a\cosh b + \cosh a \sinh b \,.  \label{eq:10}
\end{align}
When the system is defined on the circle,
\begin{equation}
  \label{eq:7}
  Z= \sum_{\sigma, \tau}e^{-H(\sigma,\tau)} = (\cosh q)^L\prod_i \cosh J_i
  + (\sinh q)^L\prod_i \sinh J_i \;,
\end{equation}
by~\eqref{eq:9} and~\eqref{eq:10}, and by the parity of the 
hyperbolic sine and cosine.
When the system is defined on the segment
\begin{equation}
  \label{eq:8}
  Z= \sum_{\sigma, \tau}e^{-H(\sigma,\tau)} = (\cosh q)^L\prod_i \cosh J_i \,.
\end{equation}
In this simple $1$D case it is also easy to exhibit a function $g$
such that ${\cal J}(g)\ne 0$.
Let us take for instance the family $g_i(\sigma,\tau)=\sigma_i\tau_{i+1}$.
For the Ising system on the segment,
\begin{align}
  {\cal J}(g_i) =&\langle \sigma_i\tau_{i+1}- \tau_i\sigma_{i+1}\rangle=Z^{-1}
  \sum_{\sigma, \tau}e^{-H(\sigma,\tau)}
  (\sigma_i\tau_{i+1}- \tau_i\sigma_{i+1})
  \nonumber\\
  =&Z^{-1}2^L
  \sum_{\sigma}\left[\sigma_i\sinh(J_i\sigma_i+q\sigma_{i+1})\prod_{l\ne
      i}\cosh(J_{l}\sigma_{l}+q\sigma_{l+1})\right. \nonumber\\
  &\left. - \sigma_{i+1}
    \sinh(J_{i-1}\sigma_{i-1}+q\sigma_{i})\prod_{l\ne
      i}\cosh(J_{l-1}\sigma_{l-1}+q\sigma_{l})\right] \nonumber\\
  =&\tanh J_i(1-\tanh^2q) \,. \label{curr1D}
\end{align}

The second option to generalise the \ac{PCA} dynamics defined by the Hamiltonian~\eqref{ising1Drev}
is to define a partially asymmetric system. 
The proof of the \ac{WBC}~\eqref{eq:weakreversibility1} relies on the
hypothesis that the system is translationally invariant.
As such, the system has to be defined on the
circle with $J_i=J$ for all $1\leq i\leq L$. The pair Hamiltonian then becomes
 \begin{equation}
    \label{ising1Dirrev2}
    H(\sigma,\tau) = -\sum_{i=1}^L \left[ \alpha
      J\,\sigma_{i}\,\tau_{i+1}+(1-\alpha)J\tau_{i}\,\sigma_{i+1}
      + q\sigma_i\tau_i\right] \;,
  \end{equation}
where $0<\alpha<1$.
The proof of \ac{WBC} for the $2$D version of~\eqref{ising1Dirrev} is presented in
Section~\ref{sec:global-balance-2d}.
The actual proof of the \ac{WBC} for~\eqref{ising1Dirrev2} is similar with respect to that -- but much
easier. It is left as an exercise to the reader.

\subsection{Weak balance for $2$D nearest neighbour completely
  asymmetric Ising systems}
\label{sec:global-balance-2d}

We now consider the translationally invariant
Ising model on a $2$D square lattice $\Lambda$ with $N\times N$ sites, with
periodic boundary conditions. 
To this purpose we introduce some notation. 
The configuration $\sigma\in\{-1,+1\}^\Lambda$ has value
$\sigma_{i,j}$ in the site $x=(i,j)$ of the torus $\Lambda$.    
The  Hamiltonian of the system is defined by the expression
\begin{equation}
    \label{eq:dss3}
    H(\sigma) = -\sum_{\substack{\{(i,j),(k,l)\}\\\text{n.n.\ pairs}}}
    J\,\sigma_{i,j}\,\sigma_{k,l} \;,
  \end{equation}
that we can rewrite more explicitly as
 \begin{equation}
   \label{ising} 
   H(\sigma) = -J\,\sum_{i,j=1}^N \big(\sigma_{i,j}\,\sigma_{i,j+1} +
   \sigma_{ij}\,\sigma_{i+1,j}\big) \,.
\end{equation}    
Similarly to the previous section, the reversible \ac{PCA} dynamics 
associated to~\eqref{ising} is determined by the pair Hamiltonian 
\begin{align}
  H(\sigma,\tau) =& -\frac{J}{2}\,\sum_{i,j=1}^N \big(\sigma_{i,j}\,\tau_{i,j+1} + \sigma_{i,j+1}\,\tau_{i,j}
    + \sigma_{i+1,j}\,\tau_{i,j} + \sigma_{ij}\,\tau_{i+1,j}\big)\nonumber\\
    &- q\, \sum_{i,j=1}^N\sigma_{i,j}\,\tau_{i,j} \,. \label{revers}
    \end{align}
According to~\eqref{eq:6} and~\eqref{eq:11}, the Hamiltonian above defines an homogeneous 
\ac{MC},
which is again ergodic and reversible with respect to the measure
$\pi(\sigma)$, given by formulas~\eqref{eq:stationary1} and~\eqref{revers}.

To generalise this picture to irreversible \acp{MC},
we define an irreversible \ac{PCA} considering a 
completely asymmetric version of the dynamics above. 
We define the following Hamiltonian 
\begin{equation}\label{irrev}
      H(\sigma,\tau) = -J\,\sum_{i,j=1}^N (\sigma_{i,j}\,\tau_{i,j+1} +
      \sigma_{i,j}\,\tau_{i+1,j}) - q\, \sum_{i,j=1}^N
      \sigma_{i,j}\,\tau_{i,j}
\end{equation}

\begin{prp}
        The irreversible \ac{PCA} defined as in~\eqref{eq:6}--\eqref{eq:11} by the
        Hamiltonian~\eqref{irrev} is weakly balanced, i.e.,
        \begin{equation}
          \label{eq:weakreversibility}
          \sum_{\tau}e^{-H(\sigma,\tau)} =
          \sum_{\tau}e^{-H(\tau,\sigma)} \;, \quad 
          \forall \; \sigma\in{\cal X} \,.
        \end{equation}
        Hence, $\pi(\sigma)= \frac{Z_\sigma}{Z}$ is the
        stationary measure of such an
        irreversible \ac{PCA}. 
\end{prp}

\begin{proof}
  By direct computation we have that
  \begin{align}\label{right}
    \sum_{\tau}e^{-H(\sigma,\tau)} &=: Z_\sigma^\rightarrow =
    2^{|\Lambda|}\prod_{i,j}\cosh
    \left(J(\sigma_{i-1,j}+\sigma_{i,j-1})+q\sigma_{i,j}\right) \;, \\
    \label{left}
    \sum_{\tau}e^{-H(\tau,\sigma)} &=: Z_\sigma^\leftarrow =
    2^{|\Lambda|}\prod_{i,j}\cosh \left(J(\sigma_{i+1,j}+\sigma_{i,j+1})+q\sigma_{i,j}\right) \,.
\end{align}
The parity of the hyperbolic
cosine then yields
\begin{equation}\label{values}
  \cosh \left(J(\sigma_{i\pm 1,j}+\sigma_{i,j\pm 1})+q\sigma_{i,j}\right) =
  \begin{cases}
    \cosh(2J+q)& \text{if } \sigma_{i\pm 1,j}=\sigma_{i,j\pm 1}=\sigma_{i,j} \;,\\
    \cosh(2J-q)& \text{if } \sigma_{i\pm 1,j}=\sigma_{i,j\pm
      1}=-\sigma_{i,j} \;,\\
    \cosh(q)& \text{if } \sigma_{i\pm 1,j}\ne \sigma_{i,j\pm 1} \,.
  \end{cases}
\end{equation}
The three different values depend on the configuration $\sigma$ when
it is represented in terms of Peierls contours.  Indeed, once a
configuration $\sigma$ and the corresponding set of Peierls contour
$\Gamma$ are chosen, then the contribution of the single term $i,j$
appearing in the products of~\eqref{right} (resp.~\eqref{left}) depends on the
contribution the dual edges below and to the left (resp. above and to
the right) of the site $i,j$. If none of them belongs to any
Peierls contour, the contribution will be $\cosh(2J+q)$; if both of
them belong to some Peierls contour, then the contribution will be
$\cosh(2J-q)$; if exactly one of them belongs to a Peierls contour and the
other does not, then the contribution will be $\cosh(q)$.  
Hence, the first part of the Proposition follows simply noting that for all
configurations of Peierls contours the number of elbows in the
SW direction is the same of the number of elbows NE-oriented. 
\end{proof}

\begin{rmk}
  Remark~\ref{rmk:beethoven_no5} implies that Theorem~\ref{thm-dss} holds
  also in this case. Thus, if $e^{-2q} =
  o\left(|\Lambda|^{\frac{1}{2}}\right)$ as $|\Lambda| \to \infty$
  then $\pi(\sigma)$, defined by~\eqref{eq:stationary1}
  and~\eqref{irrev}, tends asymptotically in total-variation distance
  to the Gibbs measure $\pi_G(\sigma)$, defined by~\eqref{gibbs} and~\eqref{ising}.
\end{rmk}

\section{Parallelization, phase transition,
  Ising waves}
\label{sec:numer-results:-ising}

In this section we present some numerical studies about the $2$D Ising
model \ac{PCA} we have discussed in
Section~\ref{sec:global-balance-2d} above, that is, the 
discrete-time \ac{MC} defined by~\eqref{eq:6}--\eqref{eq:11}.
Both~\eqref{revers} and~\eqref{irrev} will be considered as the
Hamiltonian of the system.

\acp{MC} can be easily implemented and simulated on any computer using
the so-called \emph{random mapping representation}, see e.g.~\cite{haggstrom2002finite}.
Random processes of the kind of \ac{PCA} can be simulated using
the very same approach.
The product form of the transition
matrix~\eqref{eq:1} ensures that the spins flip independently, so 
a unique random map will serve all the spins.
To fix the ideas, let us consider the reversible \ac{PCA} defined
by~\eqref{eq:6}, \eqref{eq:11} and~\eqref{revers}.
If we define the local field to be
\begin{equation}
  \label{eq:9a}
  h_{i,j}(\sigma) = \frac{J}{2} \left( \sigma_{i-1,j} +
     \sigma_{i+1,j} + \sigma_{i,j-1} + \sigma_{i, j+1} \right)\:.
\end{equation}
Then equation~\eqref{revers} becomes
\begin{equation}
  \label{eq:10a}
  H(\sigma,\tau) = -\sum_{i,j=1}^N \left[h_{i,j}(\sigma) + q
    \,\sigma_{i,j} \right]\,\tau_{i,j} \:.
\end{equation}
Thus, the transition
probabilities~\eqref{eq:6}--\eqref{eq:11} become
\begin{equation}
  \label{eq:13a}
  P(\sigma,\tau) = \prod_{i,j} \frac{\exp
    \left[\tau_{i,j}\left(h_{i,j}(\sigma) +
        q\sigma_{i,j}\right)\right]}{2\cosh \left(h_{i,j}(\sigma) +
      q\sigma_{i,j}\right)} \:.
\end{equation}
The update procedure of a spin at vertex $(i,j)$ is
sketched in the pseudocode given by Listing~\ref{lst:meta}.
\begin{lstlisting}[float=!bp,caption=Update procedure for the spin in site \textsl{(i,j)}, label=lst:meta]
  for each neighbour (k,l) of (i,j)
      p += sigma((k,l))
  end for
  p *= -J
  p -= q * sigma((i,j))
  p = 1 / (1 + exp(2p))

  if rand() < p then
      tau((i,j)) = 1
  else
      tau((i,j)) = -1
  end if
\end{lstlisting}
The number of floating point operations required
for the update of a single spin is very low. Indeed, the problem of
simulating the evolution of such a \ac{MC} is not particularly complex,
involving only simple operations on many data elements.
This is particularly suited for a \ac{GPU} because it closely
resembles the operations involved in graphic applications.
Moreover, from~\eqref{eq:1} there is no dependency in
the probabilistic update rule for the evolution of the spins. 
Therefore they can be updated
simultaneously, making the problem embarrassingly parallel.
The natural choice to efficiently implement such a model seems to be \ac{CUDA}.

\ac{CUDA} maps geometry onto blocks of threads executing
simultaneously.
It is then natural to map a configuration $\sigma$, stored as a
square matrix of size $N\times N$, onto a square grid of blocks.
Each block is able to run multiple threads, simultaneously updating a
portion of the configuration matrix $\sigma$.
The update rule~\eqref{eq:1} ensures that no inter-thread dependencies
arise as the transition probabilities depend only
on the previous time-step configuration.
The only aspect that must be carefully regarded is the memory access,
because the memory is the main potential performance inhibitor in \acp{GPU}.
It is hence very important to optimise for memory access.
Dividing the two dimensional matrix into a set of square tiles is a
good strategy to efficiently use the \ac{GPU}'s shared memory and minimise the
number of global memory reads.
However, the presence of periodic boundary conditions causes a non
uniform access pattern on the boundary, that is,
the impossibility to achieve coalesced memory and so high memory
bandwidth.
This latter issue disappears if one considers non-periodic boundary
conditions.

\begin{figure}[!htbp]
\centering
\includegraphics[width=0.75\textwidth]{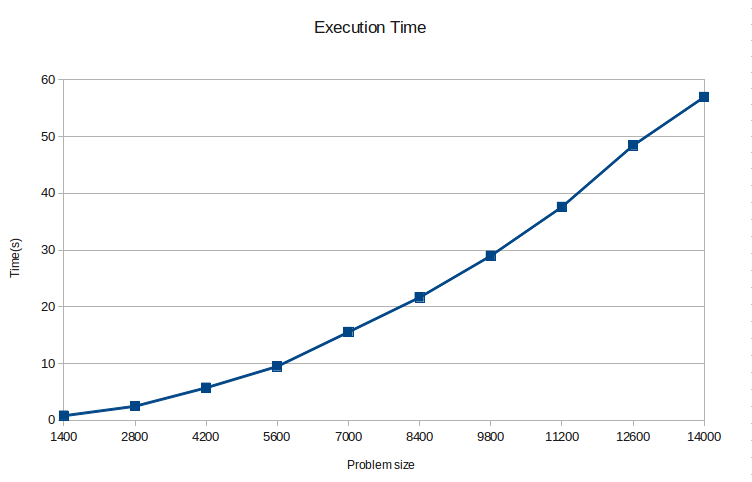}
\caption{Execution time of 500 steps of the \ac{PCA} 
    dynamics defined by~\eqref{eq:6},~\eqref{eq:11}
    and~\eqref{revers}. Different values of $N$ are considered.
    Simulations were run on a Nvidia Tesla K$20$s
    \ac{GPU}.} 
\label{res}
\end{figure}

Figure~\ref{res} shows the execution time of 500 successive
transitions of the \ac{PCA} for different values of the number of
sites $N$ in each dimension. In particular, we see that for a square
lattice of size $14000\times14000$ the execution time is less than one
minute.
Together with Theorem~\ref{thm-dss} this paves the way for an
extremely sped-up Gibbs sampling.

\begin{figure}[!tbp]
  \centering
  \subfloat[][$q=1.0$]{
    \includegraphics[width=0.65\textwidth]{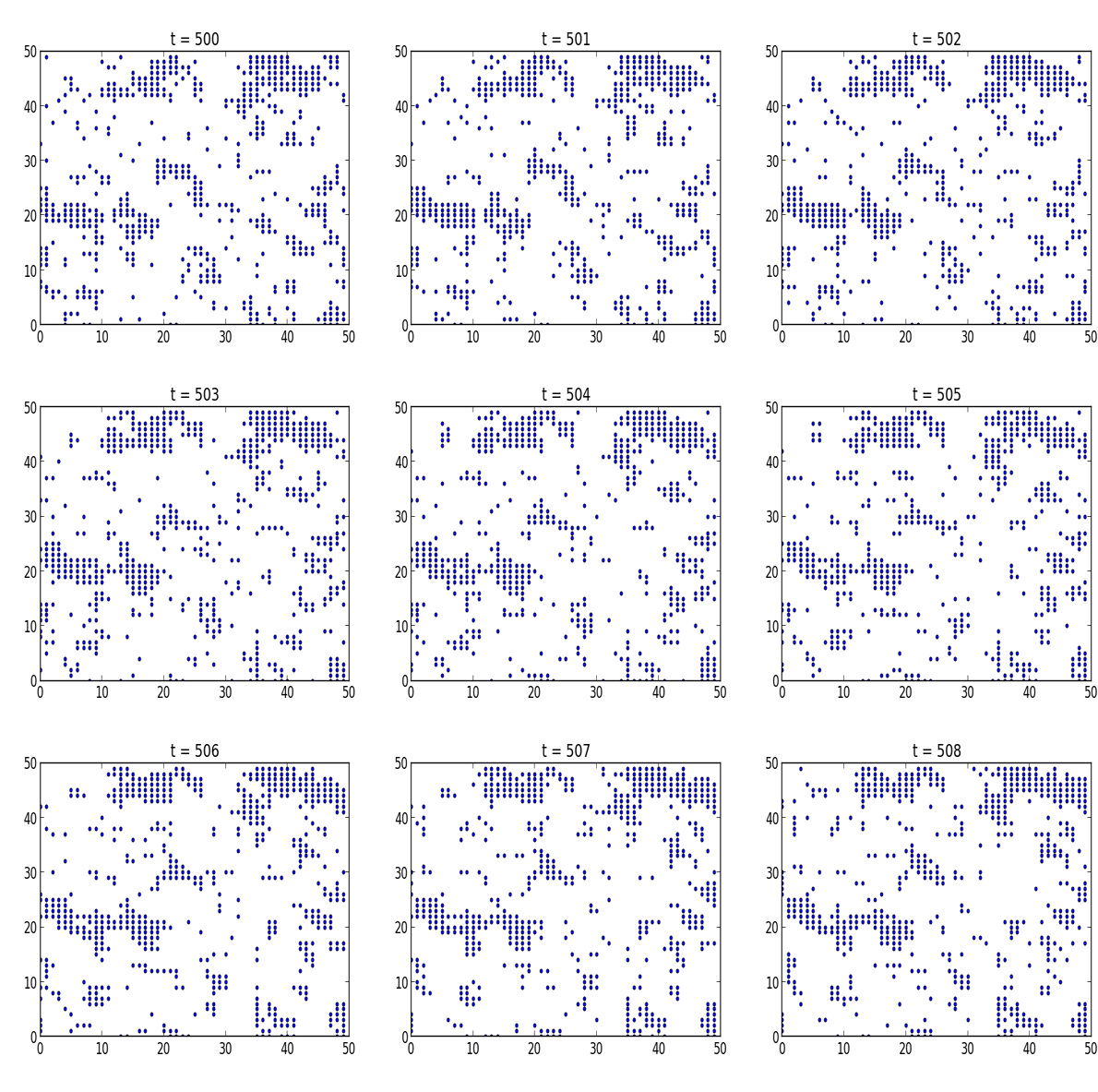}
    \label{fig:J0.5q1.0}
  }
  \\
  \subfloat[][$q=3.0$]{
    \includegraphics[width=0.65\textwidth]{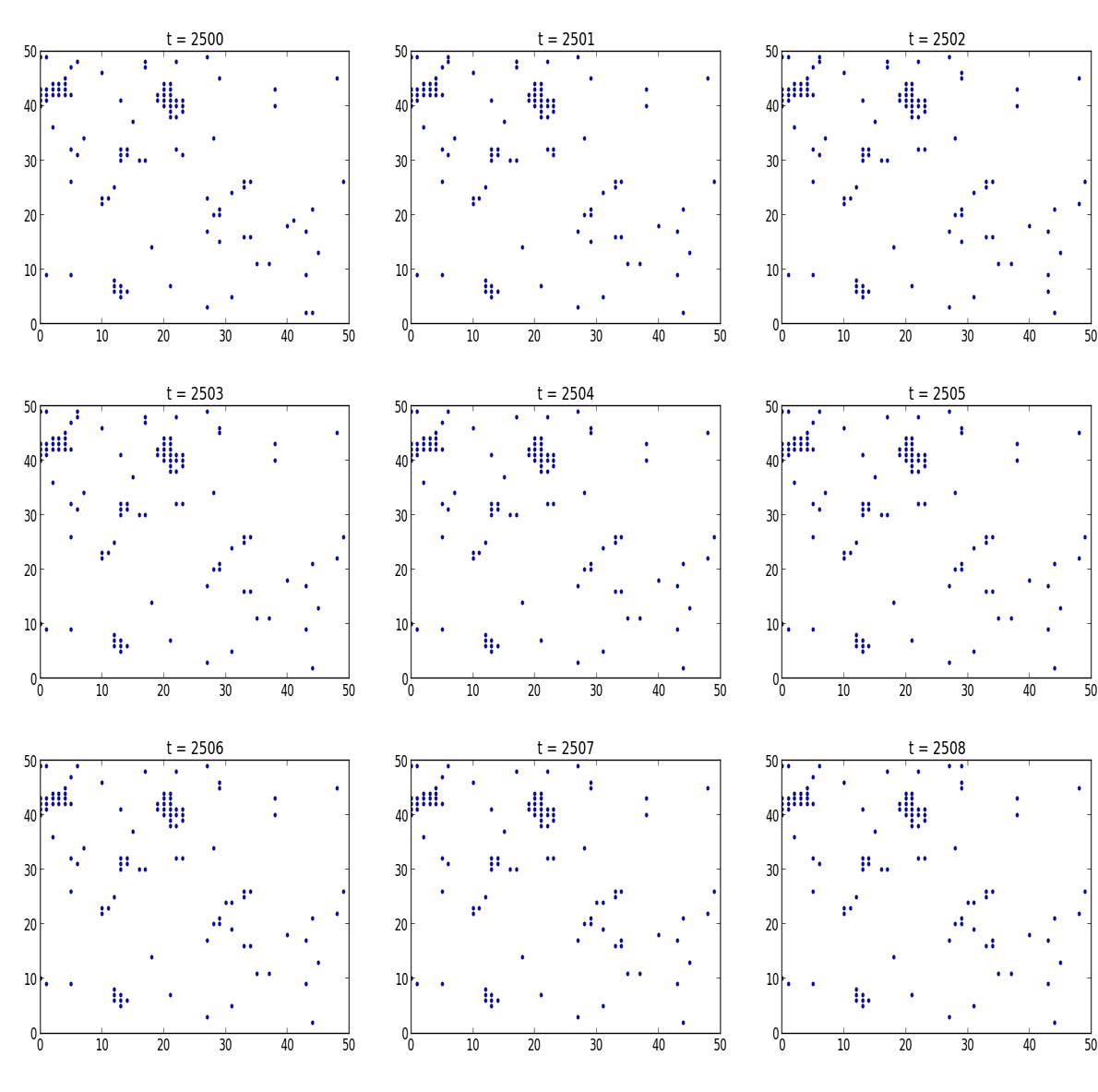}
    \label{fig:J0.5q3.0}
  }
  \caption[]{Simulation of nine successive steps of the \ac{PCA} 
    dynamics defined by~\eqref{eq:6},\eqref{eq:11} and~\eqref{irrev},
    with $N=50$, $J=0.5$ and different values of $q$.
    Blue dots represents $+1$ spins.
  }
  \label{fig:J0.5qvar}
\end{figure}

Let us now consider the irreversible \ac{PCA} defined
by~\eqref{eq:6},\eqref{eq:11} and~\eqref{irrev}, 
with initial state $\sigma_{i,j}=-1$ for all sites $(i,j)$.
To have fairly readable pictures, we set the size 
of the square lattice equal to $50\times50$.
The evolution of the dynamics for
$J=0.5$ and $q=1.0$ is presented in
Figure~\ref{fig:J0.5qvar}\subref*{fig:J0.5q1.0}, whereas the evolution
for $J=0.5$ and $q=3.0$~Figure~\ref{fig:J0.5qvar}\subref*{fig:J0.5q3.0}.
In the latter case the product
$|\Lambda|^\frac{1}{2} \, e^{-2q}$ is small, then from Theorem~\ref{thm-dss}
we may expect the typical configurations of
\acp{PCA} and serial Glauber dynamics to be rather alike. Indeed,
Figure~\ref{fig:J0.5qvar}\subref*{fig:J0.5q3.0}
displays roundish clusters of $+1$ spins, quite similar to the standard Ising
droplets.
Clearly the total asymmetry of the interaction is still visible as the
droplets present a moderate degree of stretching along the SE
direction.
The droplets stretch is even more evident in
Figure~\ref{fig:J0.5qvar}\subref*{fig:J0.5q1.0}, where the
contributions to the Hamiltonian due to the local field and to the
inertial term are comparable.

\begin{figure}[!tbp]
  \centering
  \subfloat[][$J=0.5$]{
    \includegraphics[width=0.65\textwidth]{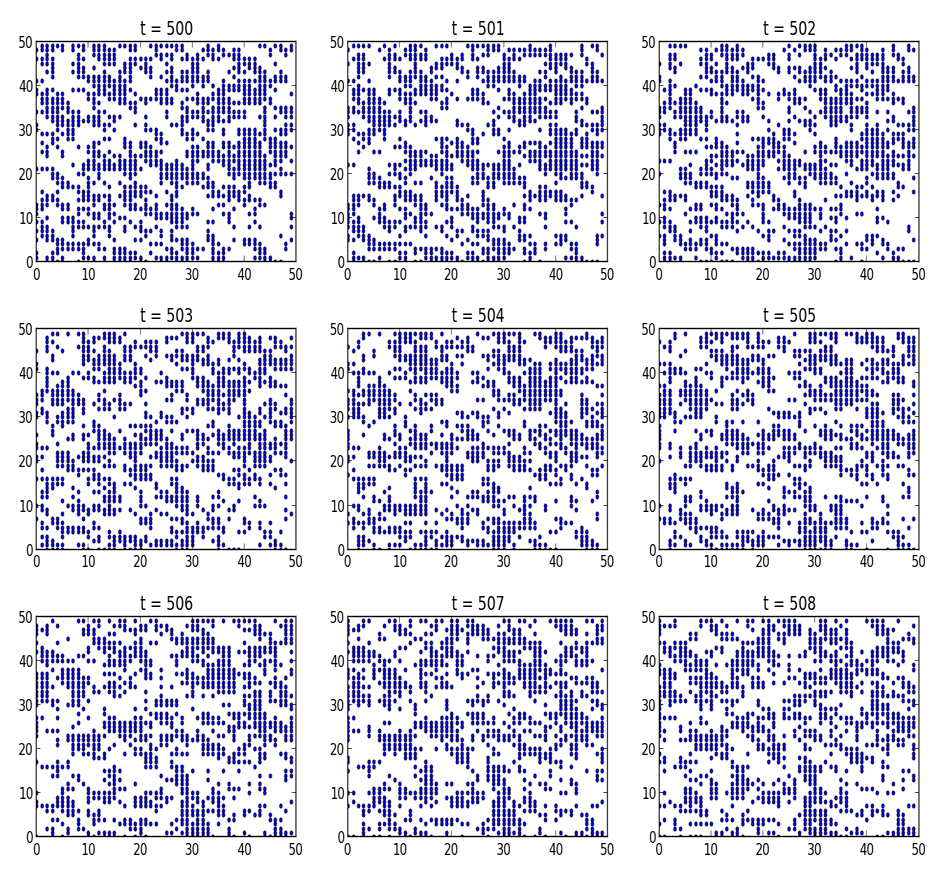}
    \label{fig:J0.5q0.5}
  }
  \\
  \subfloat[][$J=1.0$]{
    \includegraphics[width=0.65\textwidth]{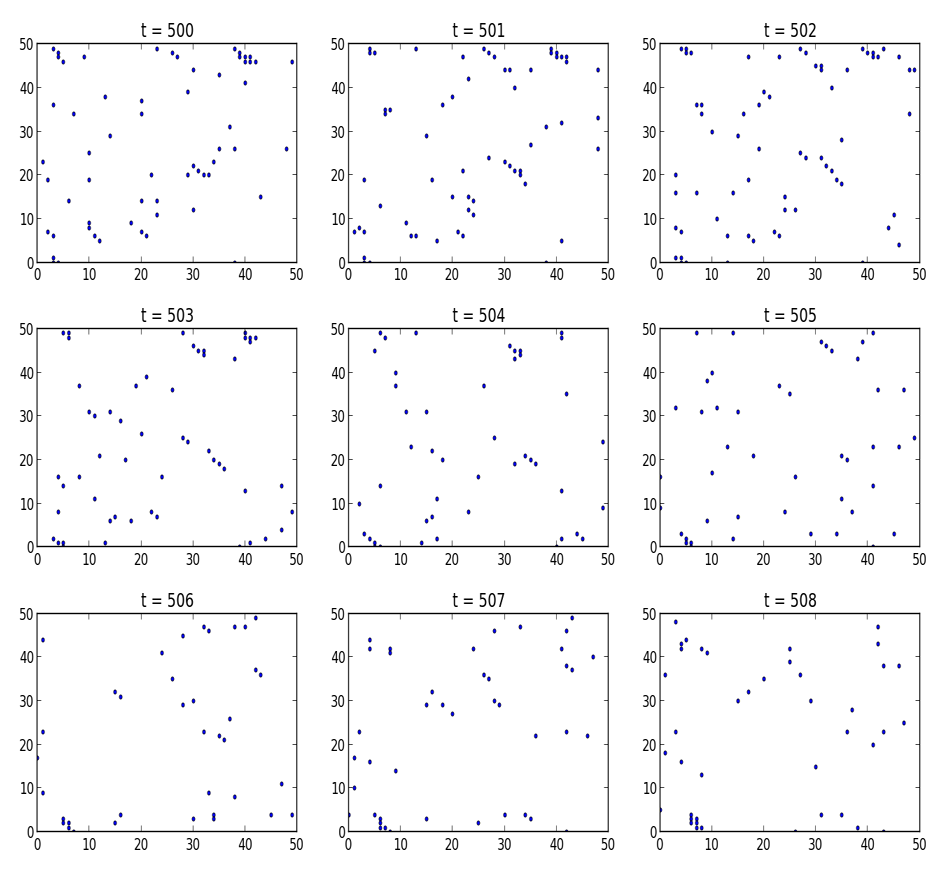}
    \label{fig:J1.0q0.5}
  }
  \caption[]{Simulation of nine successive steps of the \ac{PCA} 
    dynamics defined by~\eqref{eq:6},\eqref{eq:11} and~\eqref{irrev},
    with $N=50$, $q=0.5$ and different values of $J$. 
    Blue dots represents $+1$ spins.
    Figure~\subref{fig:J1.0q0.5} shows a negative
    spontaneous magnetisation which is not observed
    in~\subref{fig:J0.5q0.5}.}
  \label{fig:Jvarq0.5}
\end{figure}

In Figure~\ref{fig:Jvarq0.5} the value of $q$ is fixed to the value
$\frac{1}{2}$. For a large value of $J$ the system
exhibits spontaneous magnetization
(Fig.~\ref{fig:Jvarq0.5}\subref*{fig:J1.0q0.5}), whereas for a small
value of $q$ the system exhibit a paramagnetic behaviour
(Fig.~\ref{fig:Jvarq0.5}\subref*{fig:J0.5q0.5}).
Figure~\ref{fig:Jvarq0.5} then suggests the presence of a phase
transition, and a comparison of
Figure~\ref{fig:Jvarq0.5}\subref*{fig:J0.5q0.5} with
Figures~\ref{fig:J0.5qvar}\subref*{fig:J0.5q1.0}--\subref*{fig:J0.5q3.0}
clearly implies that the critical temperature, if any exists, must be
function of both $J$ and $q$.

\begin{figure}[!tbp]
  \centering
  \includegraphics[width=0.75\textwidth]{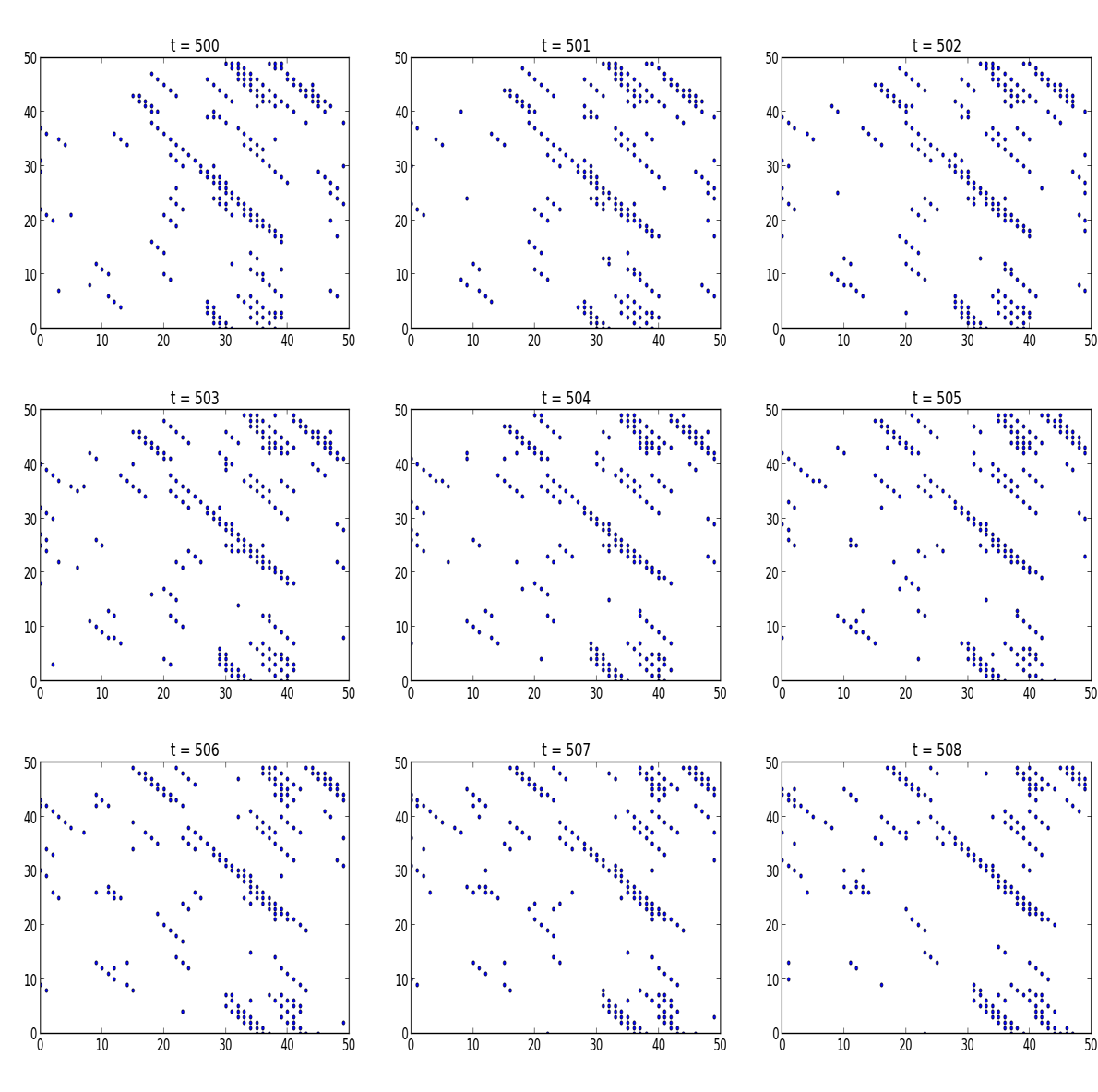}
  \caption{Simulation of nine successive steps of the \ac{PCA} 
    dynamics defined by~\eqref{eq:6},\eqref{eq:11} and~\eqref{irrev},
    with $N=50$, $J=1.5$ and $q=0.1$. Blue dots represents $+1$ spins.}
  \label{fig:J1.5q0.1}
\end{figure}

The last argument has shown the need to explore the whole plane $J,q$.
A region of this plane we find rather interesting is that 
for $J$ large and $q$ small. Here
the spin-flipping is not hampered by the contribution $-q\,
\sigma_{i,j}\,\tau_{i,j}$, so the alignment of a spin with its
souhtern and western nearest neighbour costs virtually nothing.
Figure~\ref{fig:J1.5q0.1} shows the evolution for $J=1.5$ and $q=0.1$.
Due to the mechanism we have just described, the droplets steadily
drift in the NE direction. The constant drift suggests the presence of
a current of a family of functions 
\begin{equation}
  \label{eq:12}
  g_{i,j}(\sigma, \tau) = \tau_{i,j}(\sigma_{i-1,j} + \sigma_{i, j-1}) \;,
\end{equation}
which generalise to the $2$D case what we have already discovered
in~\eqref{curr1D}.
Figure~\ref{fig:J1.5q0.1} also suggests that there is another
mechanism at work along the mentioned drift. Due to the total
asymmetry of the interaction, the growth of the droplets along the SW
direction is preferred. We propose for such elongated, drifting
droplets the name of \emph{Ising waves}.

\clearpage

    \section*{Acknowledgments}
    The first author would like to express appreciation to Alexander Agathos (West
    University of Timisoara) and
    Salvatore Filippone (University of Rome ``Tor Vergata'') for the
    useful comments regarding the \ac{CUDA} implementation;
    he would also like to address a very special thank 
    to Thomas Falch (NTNU) and Johannes Kvam (NTNU) for their
    gargantuan help in developing and running the code.

    \addtocontents{toc}{\protect\vspace{\beforebibskip}}
    \addcontentsline{toc}{section}{\refname}    
    \bibliographystyle{alpha}
    \bibliography{LS}
\end{document}